# Molecular beam epitaxy growth of non-magnetic Weyl semimetal LaAlGe thin film

Niraj Bhattarai[1,2,*], Andrew W. Forbes[1,2], Rajendra P. Dulal[3], Ian L. Pegg[1,2], and John Philip[1,2]

[1]Department of Physics, The Catholic University of America, Washington, DC, 20064, USA.
[2]The Vitreous State Laboratory, The Catholic University of America, Washington, DC, 20064, USA.
[3]Institute for Quantum Physics, Advanced Physics Laboratory, Chapman University, MD, 20866, USA.

*Corresponding author: bhattarai@cua.edu

## Abstract

Here, we report a detailed method of growing LaAlGe, a non-magnetic Weyl semimetal, thin film on silicon(100) substrates by molecular beam epitaxy and their structural and electrical characterizations. 50 nm thick LaAlGe films were deposited and annealed for 16 hours in situ at a temperature 793 K. As-grown high-quality films showed uniform surface topography and near ideal stoichiometry with a body-centered tetragonal crystal structure. Temperature-dependent longitudinal resistivity can be understood with dominant interband s-d electron-phonon scattering in the temperature range 5 − 40 K. Hall measurements confirmed the semimetallic nature of the films with electron dominated charge carrier density $\sim 7.15 \times 10^{21}$ cm$^{-3}$ at 5 K.

## INTRODUCTION

Recent theoretical and experimental study has shown that the family of $R$AlGe ($R$ = La, Ce, Pr) are Weyl semimetals that offer remarkable tunability including both magnetic and non-magnetic compounds as well as both type I or type II Weyl Semimetal (WSM) states [1, 2, 3, 4]. Weyl semimetals are an exciting class of materials in which the low energy electronic excitations − massless Weyl fermions − disperse linearly along all three momentum directions through the Weyl nodes. This hallmark feature is observed in the materials' electronic spectra, precipitating a wide variety of useful properties that hold promise for both fundamental and technological applications such as topological qubits, low-power electronics, and spintronics [5, 6, 7]. The origin of WSM behavior in materials lies in either breaking crystal inversion symmetry, time-reversal symmetry, or both [8, 9, 10]. Breaking at least one of these symmetries can induce unusual physical phenomena, such as spin current without a simultaneous charge current, negative magnetoresistance, Fermi arcs, anomalous quantum Hall effect, and chiral magnetic effects [11, 12, 13, 14, 15]. Hence, it is of great interest to enable a wide range of experimental studies on this class of materials, and the first step is to establish the details for the growth of device applicable thin film structures and their structural and electrical characterizations.

LaAlGe is a non-magnetic material relatively understudied compared to its isostructural sister compounds− CeAlGe and PrAlGe, and there are only a few reports in literature. Back in 1991, Guloy and Corbett [16] first synthesized and showed that LaAlGe crystallizes in a body-centered tetragonal Bravais lattice of type LaPtSi, with an inversion symmetry − breaking space-group $I4_1md$ (109) with lattice parameters $a = b = 4.33$ Å and $c = 14.82$ Å. Later, a few other reports [3, 17] have confirmed this crystal structure. Recently, from the theoretical calculations of low energy band structure [1, 16], and experimental study based on photoemission data [3], LaAlGe has been realized as a space-inversion symmetry breaking type II WSM with tilted Weyl points arising at the touching of electron and hole pockets, different from ordinary Weyl points with a point-like Fermi surface in type I WSMs. These reports are on the growth and study of bulk single crystals by flux method, and to the best of our knowledge, there have been no reports on thin film LaAlGe to date. As a rare-earth-based compound, LaAlGe is promising for a variety of useful devices that can be fabricated using thin films. Thin film structures can give rise to unusual phenomena that have never been observed in their bulk counterparts [18].

In this article, we report the growth and characterization of 50 nm thick LaAlGe thin films grown by molecular beam epitaxy (MBE). We have grown polycrystalline, large-area LaAlGe thin film samples on Si (100) wafers and studied the structural and morphological characteristics. Analyses show we obtained near ideal stoichiometric thin films with a body-centered tetragonal crystal structure. We have also investigated the transport characteristics such as longitudinal resistivity and Hall effect measurements of these films, which confirms they are semimetallic and display a charge transport dominated by electrons.

## EXPERIMENTAL

High-quality LaAlGe thin film samples were grown by MBE deposition on semi-insulating silicon (100) substrates under a vacuum condition better than $1.1 \times 10^{-9}$ Torr during deposition. Before deposition, the silicon wafers were cut into the desired shape and size and were sonicated in a hot soap-water solution for 5 minutes. Then the substrates were transferred into a clean beaker and rinsed several times using acetone, isopropyl alcohol, and deionized water successively. They were then etched in 20% hydrofluoric acid solution for one minute to remove the native oxide. Thus prepared substrates were mounted immediately on a temperature-controlled heater plate and loaded into an MBE chamber. Chunks of lanthanum (99.99% metal basis, Sigma Aldrich, St. Louis, Missouri), initially kept in an oil bath were cleaned using the standard cleaning protocol for cleaning deposition materials shipped in oil and loaded into the high vacuum in a tungsten crucible. Pellets of aluminum (99.99% metal basis, Sigma Aldrich) and germanium (99.99% metal basis, Sigma Aldrich) were used for the e-beam evaporation. The silicon substrates were preheated at 523 K for 30 minutes. A 10 nm buffer layer of MgO was deposited followed by e-beam evaporation of stoichiometric amounts of lanthanum, aluminum, and germanium, keeping the substrate temperature at 623 K throughout the deposition. Film thickness and uniformity of the deposition was monitored by quartz crystal rate monitors. After completion of the deposition, the substrate temperature was ramped up to 793 K and held for 16 hours to grow the LaAlGe thin films *in situ*. After the film was cooled down to room temperature, a protective layer of 5 nm MgO was deposited to reduce oxidation of the film.

A Thermo/ARL X'TRA diffractometer (Thermo Fisher Scientific, Waltham, Massachusetts) with (Cu-K$_\alpha$) radiation (1.54 Å) in the theta-theta Bragg-Brentano geometry was used to collect the X-ray diffraction patterns to confirm the phase of the thin films; structural parameters were estimated by a full Rietveld refinement using JADE 9 software (Materials Data Inc., Livermore, California); the morphology was examined by scanning electron microscopy (SEM) model JEOL JSM-

5910LV (JEOL USA Inc., Peabody, Massachusetts) and contact atomic force microscopy (AFM) (Alpha300 RA, WITec, Ulm, Germany). Similarly, the composition was analyzed by energy dispersive X-ray spectroscopy (EDS) analysis using Oxford instruments Ultim Max (Oxford Instruments, High Wycombe, U.K.) with 170 mm$^2$ aperture coupled with SEM [JEOL JSM-5910LV].

Electrical measurements were carried out on these thin films connected with gold wire attached with indium metal electrodes. The electrical transport measurements were carried out in standard four-probe geometry, whereas Hall measurements were done using a Hall shape device grown by metal contact mask during the film deposition. The measurements were carried out in a Quantum Design Physical Property Measurement System with the AC transport option. We will discuss results based on the longitudinal resistivity and Hall measurements of these films.

## RESULTS AND ANALYSIS

Figure 1(a) displays the X-ray diffraction (XRD) pattern of 50 nm thin film of LaAlGe deposited on silicon (100) substrate by MBE method. All of the observed XRD peaks can be indexed based on a body-centered tetragonal Bravais lattice with space group I4$_1$md (109), whose atomic arrangement is shown in Figure 1(b). These are the strongest reflections for this crystal structure, as seen in the simulated pattern represented by the blue curve in Figure 1(a). The pattern is simulated using the lattice parameters taken from reference [17].

The Scherrer equation; $D = \frac{K\lambda}{\xi\cos(\theta)}$; is used to calculate the average crystallite size (*D*), which is inversely related to the full-width-half-maximum (ξ) of the individual peaks at position θ (half of 2θ). Here, *K* is a Scherrer constant and λ is the wavelength of the x-rays used, 1.54 Å [19]. The average crystallite size calculated from the strongest peak (112) is found to be 181 Å. Table I summarizes the estimate of the average crystallite size corresponding to each diffraction peak calculated using experimental data.

To obtain the best estimates of the structural parameters, we have carried out a whole pattern fitting Rietveld refinement of the x-ray diffraction peaks in the range 10° ≤ 2θ ≤ 65° (we have excluded the silicon background by choosing this range). The refinement was initiated using the lattice parameters in the literature [16]. The refinement converged on the unit cell parameters: *a* = *b* = 4.32 (±0.005) Å, and *c* =14.80 (±0.017) Å and the best fit was obtained with goodness of fit value $(G) = (\frac{R_{wp}}{R_{exp}}) = 1.233$; where $R_{wp}$ is the *R*-factor = 10.33%, and $R_{exp}$ is called the expected *R*-factor = 8.38%. The black curve represents the best fit in Figure 1(a). With all the atoms in the 4a Wyckoff sites, the atomic position and occupancy for all atoms in LaAlGe are given in Table II. These outcomes are in good agreement with the experimental results reported in the literature [16, 17]. The crystallite size obtained from the whole pattern fit is 234 Å. This value is comparable to the estimates from individual peaks.

The atomic structure shown in Figure 1 (b) consists of lanthanum, aluminum, and germanium atoms represented by red, blue, and green spheres, respectively, stacked along the c-axis forming a noncentrosymmetric structure.

The SEM image in Figure 2 (a) shows morphology of 50 nm LaAlGe thin film sample grown by MBE. We observed continuous and very uniform thin films distributed over the silicon substrate. The topography of the thin film surface obtained by AFM is shown in Figure 2 (b), which is consistent across different parts of the substrate and confirms the continuity and smoothness of the films observed from SEM images. The topography shows a variation on the same scale as the average crystallite size (~21 nm), indicating that the structures here correspond to individual crystallites.

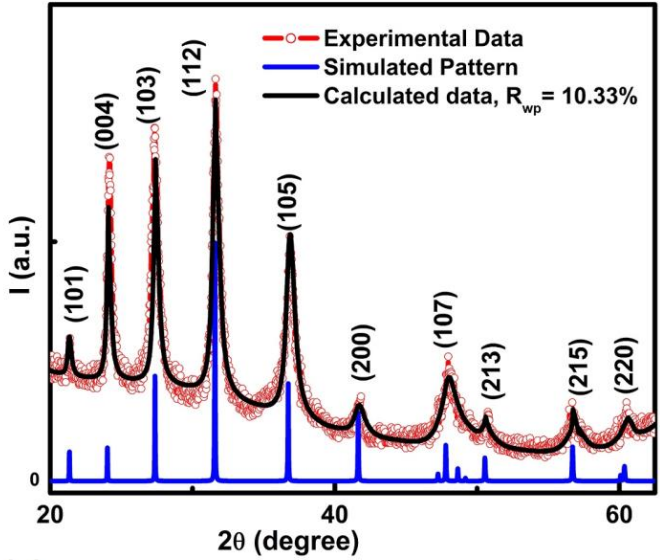 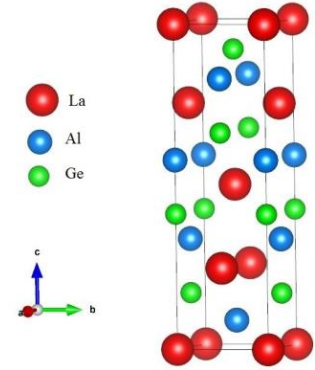

(a)  (b)

**Figure 1.** Crystallographic analysis of LaAlGe thin films: (a) The X-ray diffraction pattern: the red curve represents the experimentally observed data obtained from the X-ray diffraction measurements, the blue curve is simulated pattern, and the black curve is calculated using Rietveld refinement with a discrepancy index of $R_{wp}$ = 10.33%. (b) The $I4_1md$ (109) lattice structure of LaAlGe made using VESTA [29].

**TABLE I.** Summary of different parameters taken from individual XRD peaks. The table consists of diffraction peak position (2θ) and corresponding Miller index (h k l), interplanar spacing (d), peak full-width at half-maximum (ξ), and an estimate of average crystallite size (*D*).

| 2θ (°) | (h k l) | d (Å) | ξ (°) | *D* (Å) |
|---|---|---|---|---|
| 21.3 | (101) | 4.16 | 0.294 | 275 |
| 24.0 | (004) | 3.71 | 0.273 | 297 |
| 27.4 | (103) | 3.26 | 0.372 | 219 |
| 31.6 | (112) | 2.83 | 0.455 | 181 |
| 36.7 | (105) | 2.45 | 0.633 | 132 |
| 41.7 | (200) | 2.17 | 0.592 | 143 |
| 48.5 | (107) | 1.90 | 0.840 | 104 |
| 50.5 | (213) | 1.80 | 0.312 | 281 |
| 56.8 | (215) | 1.62 | 0.311 | 291 |
| 60.4 | (220) | 1.53 | 0.525 | 176 |

**TABLE II.** Positional parameters of atoms in LaAlGe and occupancy (Occ) with important distances in fractional coordinates.

| Atom | x | y | z | Occ |
|---|---|---|---|---|
| La | 0 | 0 | 0 | 1 |
| Al | 0 | 0 | 0.582 | 1 |
| Ge | 0 | 0 | 0.416 | 1 |

Furthermore, we have investigated the quantitative elemental analysis of the thin film samples by EDS analysis.

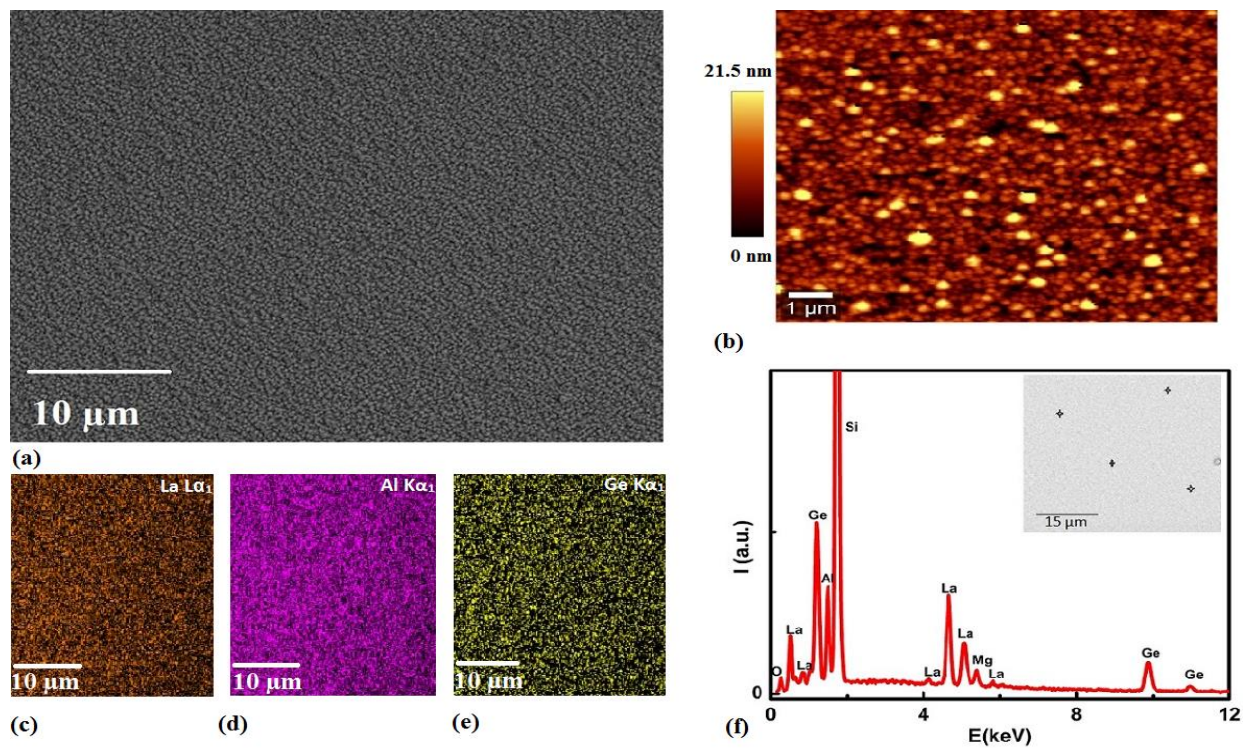

**Figure 2.** Morphology of LaAlGe thin films: (a) The SEM image. (b) Atomic force microscopy image taken by ac-contact mode over an area of 10 × 10 μm². It shows a uniform surface topography. (c), (d), and (e) EDS mapping reveals the homogenous elemental distribution (La, Al, and Ge) on the substrates. (f) Chemical mapping peaks obtained from the energy dispersive spectroscopy analysis of the film and the corresponding image used for EDS measurements as an inset in the upper right corner.

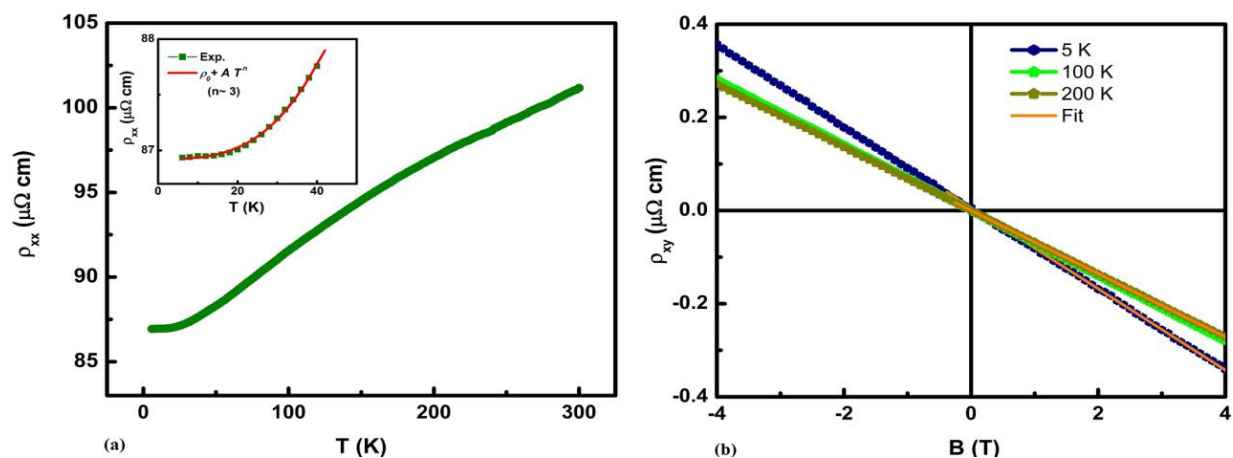

**Figure 3.** Transport measurements: (a) Temperature dependence of longitudinal resistivity of the 50 nm thin film of LaAlGe measured between 5 and 300 K, shows a metal-like behavior. Inset shows a fit using the formula $\rho_{xx} = \rho_0 + AT^n$. (b) Hall resistivity measurements taken at 5, 100, and 200 K are fitted linearly.

Figures 2 (c), (d), and (e), respectively, are the chemical mapping images taken from $L\alpha_1$, $K\alpha_1$, and $K\alpha_1$ radiation analysis, showing the homogenous distribution of La, Al, and Ge atoms on the silicon substrate. Figure 2(f) shows a typical EDS spectrum obtained from the sample, and the corresponding image is in the inset in the upper right corner. Several thin film samples were grown and the EDS results are consistent in each case. Based on EDS measurements involving multiple spot scanning on the thin film surface (results presented here are based on scanning 4 spots that are marked on the image in inset of Figure 2 (f)), we have confirmed that the lanthanum aluminum germanium were uniformly distributed over the film. The estimated average atomic (%) ratio of La:Al:Ge is equal to 34.2:32:33.8, with the standard deviation of atomic (%) ratio by 0.62, 0.98, and 1 for La, Al, and Ge respectively. The calculated ratio is very close to the expected stoichiometry of 1:1:1 for La, Al, and Ge in LaAlGe.

Figure 3 (a) shows the temperature dependence of the longitudinal resistivity ($\rho_{xx}$) from 5 to 300 K. At 5 K, the resistivity is 86.84 µΩ cm, which increases to 101.25 µΩ cm at 300 K, showing a metal-like behavior. There are no previous reports on thin films LaAlGe to compare the longitudinal resistivity; nevertheless, resistivity reported at room temperature for its single crystal is ~ 68 µΩ cm [17], which is close to our observation. The estimated residual resistivity ratio (RRR) $\rho_{xx}$(300 K)/$\rho_{xx}$(5 K) is 1.17 which is comparatively low even for semimetal [20, 21], however it is very close to that reported for flux-grown single-crystals LaAlGe (~2) [17]. It is also comparable with its isostructural family members, for example, the RRR values of bulk single crystals CeAlGe [17] and PrAlGe [4], grown by flux-method, has been reported ~2. Likewise, another study based on single crystals CeAlGe grown by floating-zone technique has reported 1.3 [2]. From these comparisons we see low RRR value could be just a feature of *R*AlGe family.

Furthermore, the inset in the upper left corner displays a fit to the resistivity data in between 5 and 40 K. The solid line in red is the fit using the formula $\rho_{xx} = \rho_0 + AT^n$, with $\rho_0$ = 87.23 µΩ cm, $A$ = 9.84 × 10$^{-6}$ µΩ cm K$^{-3}$ and $n$ = 3.08. Different values of $n$ imply different modes of scattering mechanisms that are dominant in a material. For example, $n$ = 2 is when electron–electron scattering is dominant, and n = 5 for the conventional electron-phonon scattering process according to the Bloch-Grüneisen theory [22]. But the value of $n$ (~3) in our case is different from these two scattering mechanisms. Such a mode of scattering should arise from the dominant interband s–d electron–phonon scattering, rather than the intraband s–s electron–phonon scattering at low temperature [23]. Recent study on bulk single crystals of PrAlGe [27], has also reported a similar type of T$^3$ dependent resistivity curve at low temperature regime. Also, other lanthanum based semimetallic compounds like LaBi [24] and LaSb [25], their $n$ dependence is equal to 2.99 and 4 respectively, and yttrium metal [26], as well as other transition metal carbides [23], has been reported to exhibit comparable behavior.

We investigated the Hall resistivity measurements at 5, 100, and 200 K, as shown in Figure 3 (b). The Hall resistivity measurements were carried out in a magnetic field (*B*) between ±4 T. Performing Hall measurements in positive and negative field directions helps to improve the precision of data. At a particular temperature, with increasing magnetic field strength, the observed Hall resistivity decreased monotonously, presenting ordinary Hall behavior in our thin film samples. Such ordinary Hall resistivity arises due to the Lorentz force acting on the charge carriers in the presence of an external magnetic field. The Hall resistivity, which remained unsaturated within the field range probed, is linearly dependent on the magnetic field, suggesting a strong dominance of one kind of charge carrier. Hence, it was interpreted using single-band model, $\rho_{xy} = R_0 B$. Here, $R_0 = 1/(n e)$ is the Hall effect coefficient, where $n$ is the effective charge carrier density, and e the electronic charge. At 5 K, $R_0 = -8.72 \times 10^{-8}$ $\Omega$ $cm$ $T^{-1}$, the negative sign indicates the dominance of electrons in charge transport, was calculated from the slope of the linear fit to the Hall resistivity data [Figure 3 (b)] yielding effective charge carrier density $n = \sim 7.15 \times 10^{21} cm^{-3}$ in our thin film samples.

Interestingly, the negligible difference in slope of Hall resistivity at 200 and 100 K indicate weak temperature dependent Hall coefficient at temperatures under study. A similar type of weak temperature dependent Hall resistivity has earlier been reported in bulk samples of its sister compound CeAlGe [17]. Also, the comparison of charge carrier density calculated at 5 K, 100 K ($\sim 8.90 \times 10^{21} cm^{-3}$), and 200 K ($\sim 9.18 \times 10^{21} cm^{-3}$) show very weak contribution from thermal effect on the creation of charge carriers with increasing temperature. The calculated carrier concentrations in our case is in the order of 100 times smaller than that of NbN films [28], ten times lower than that of copper, and comparable to that of WSM $Co_2TiGe$ film [15]. This indicates that LaAlGe film has the carrier concentration of that of a typical semimetal.

# SUMMARY AND CONCLUSIONS

In summary, we have successfully grown thin films of the non-magnetic Weyl semimetal candidate LaAlGe by MBE on a silicon substrate, and the detailed procedure is discussed. This method can be extended to grow thin film samples with the desired thickness. The structural characterization has shown that the film crystallized into body-centered tetragonal Bravais lattice with noncentrosymmetric space group $I4_1md$ (109) fulfilling one of the requirements to host Weyl fermions. The surface topography was found to be very uniform, with a variation on the same scale of estimated average crystallite size (~21 nm). Analysis of the low temperatures longitudinal resistivity suggests a dominant interband s–d electron–phonon scattering mechanism. From the Hall effect measurements, we observed that the electronic transport of these film samples is dominated by electrons. The typical RRR value of 1.17 and the charge carrier density calculated at different temperatures suggest the grown thin films to be semimetal. These results are very much consistent with previous experimental reports on $R$AlGe members [17, 27]. Our results indicate that high-quality LaAlGe films can be grown, and further improving the growth may lead to epitaxial LaAlGe films which may exhibit interesting physics. Thus, this research will provide foundation for understanding the rich physics of LaAlGe in thin film form.

# ACKNOWLEDGEMENTS

The authors want to thank The Vitreous State Laboratory for its financial support.

# REFERENCES

1. Chang, G., Singh, B., Xu, S.Y., Bian, G., Huang, S.M., Hsu, C.H., Belopolski, I., Alidoust, N., Sanchez, D.S., Zheng, H. and Lu, H.: *Magnetic and noncentrosymmetric Weyl fermion semimetals in the R AlGe family of compounds (R= rare earth). Physical Review B* **97**, 041104 (2018).

2. Puphal, P., Mielke, C., Kumar, N., Soh, Y., Shang, T., Medarde, M., White, J.S. and Pomjakushina, E.: *Bulk single-crystal growth of the theoretically predicted magnetic Weyl semimetals R AlGe (R= Pr, Ce)*. Physical Review Materials **3**, 024204(2019).

3. Xu, S.Y., Alidoust, N., Chang, G., Lu, H., Singh, B., Belopolski, I., Sanchez, D.S., Zhang, X., Bian, G., Zheng, H. and Husanu, M.A.: *Discovery of Lorentz-violating type II Weyl fermions in LaAlGe*. Science advances **3**, e1603266(2017).

4. Meng, B., Wu, H., Qiu, Y., Wang, C., Liu, Y., Xia, Z., Yuan, S., Chang, H. and Tian, Z.: *Large anomalous Hall effect in ferromagnetic Weyl semimetal candidate PrAlGe*. APL Materials **7**, 051110(2019).

5. Tokura, Y., Kawasaki, M. and Nagaosa, N.: *Emergent functions of quantum materials*. Nature Physics **13**, 1056(2017).

6. Huang, S.M., Xu, S.Y., Belopolski, I., Lee, C.C., Chang, G., Wang, B., Alidoust, N., Bian, G., Neupane, M., Zhang, C. and Jia, S.: *A Weyl Fermion semimetal with surface Fermi arcs in the transition metal monopnictide TaAs class*. Nature communications **6**, 7373(2015).

7. Arute, F., Arya, K., Babbush, R., Bacon, D., Bardin, J.C., Barends, R., Biswas, R., Boixo, S., Brandao, F.G., Buell, D.A. and Burkett, B.: *Quantum supremacy using a programmable superconducting processor*. Nature **574**, 505(2019).

8. Soluyanov, A.A., Gresch, D., Wang, Z., Wu, Q., Troyer, M., Dai, X. and Bernevig, B.A.: *Type-II Weyl semimetals*. Nature **527**, 495(2015).

9. Lv, B.Q., Weng, H.M., Fu, B.B., Wang, X.P., Miao, H., Ma, J., Richard, P., Huang, X.C., Zhao, L.X., Chen, G.F. and Fang, Z.: *Experimental discovery of Weyl semimetal TaAs*. Physical Review X **5**, 031013(2015).

10. Xu, S.Y., Belopolski, I., Alidoust, N., Neupane, M., Bian, G., Zhang, C., Sankar, R., Chang, G., Yuan, Z., Lee, C.C. and Huang, S.M.: *Discovery of a Weyl fermion semimetal and topological Fermi arcs*. Science **349**, 613(2015).

11. Wang, J., Lian, B. and Zhang, S.C., Generation of Spin Currents by Magnetic Field in $\mathcal{T}$-and $\mathcal{P}$-Broken Materials. In *SPIN*. World Scientific Publishing Company. doi.org/10.1142/S2010324719400137

12. Huang, X., Zhao, L., Long, Y., Wang, P., Chen, D., Yang, Z., Liang, H., Xue, M., Weng, H., Fang, Z. and Dai, X.: *Observation of the chiral-anomaly-induced negative magnetoresistance in 3D Weyl semimetal TaAs*. Physical Review X **5**, 031023(2015).

13. Son, D.T. and Spivak, B.Z.: *Chiral anomaly and classical negative magnetoresistance of Weyl metals*. Physical Review B **88**, 104412(2013).

14. Wan, X., Turner, A.M., Vishwanath, A. and Savrasov, S.Y.: *Topological semimetal and Fermi-arc surface states in the electronic structure of pyrochlore iridates*. Physical Review B **83**, 205101(2011).

15. Dulal, R.P., Dahal, B.R., Forbes, A., Bhattarai, N., Pegg, I.L. and Philip, J.: *Weak localization and small anomalous Hall conductivity in ferromagnetic Weyl semimetal $Co_2TiGe$*. Scientific reports **9**, 3342(2019).

16. Guloy, A.M. and Corbett, J.D.: Syntheses and structures of lanthanum germanide, LaGe$_{2-x}$, and lanthanum aluminum germanide, LaAlGe: interrelationships among the alpha.-ThSi$_2$,. alpha.-GdSi$_2$, and LaPtSi structure types. *Inorganic Chemistry* **30**, 4789(1991).
17. Hodovanets, H., Eckberg, C.J., Zavalij, P.Y., Kim, H., Lin, W.C., Zic, M., Campbell, D.J., Higgins, J.S. and Paglione, J.: *Single-crystal investigation of the proposed type-II Weyl semimetal CeAlGe. Physical Review B* **98**, 245132(2018).
18. Ohring, M.: Why are thin films different from the bulk? *In Laser-Induced Damage in Optical Materials.* International Society for Optics and Photonics **2114**, 624(1993).
19. Forbes, A.W., Dulal, R.P., Bhattarai, N., Pegg, I.L. and Philip, J.: *Experimental realization and magnetotransport properties of half-metallic Fe$_2$Si. Journal of Applied Physics* **125**, 243902(2019).
20. Bhattarai, N., Forbes, A.W., Dulal, R.P., Pegg, I.L. and Philip, J.: *Transport characteristics of type II Weyl semimetal MoTe$_2$ thin films grown by chemical vapor deposition. Journal of Materials Research* **35**, 454(2020).
21. Chen, B., Duan, X., Wang, H., Du, J., Zhou, Y., Xu, C., Zhang, Y., Zhang, L., Wei, M., Xia, Z. and Cao, C.: *Large magnetoresistance and superconductivity in α-gallium single crystals. npj Quantum Materials* **3**, 40(2018).
22. Ziman J.M.: *Electrons and Phonons,* Classics Series 2011 (Oxford: Oxford University Press)
23. Zhang, X., Xiao, Z., Lei, H., Toda, Y., Matsuishi, S., Kamiya, T., Ueda, S. and Hosono, H.: *Two-dimensional transition-metal electride Y$_2$C. Chemistry of Materials* **26**, 6638(2014).
24. Sun, S., Wang, Q., Guo, P.J., Liu, K. and Lei, H.: *Large magnetoresistance in LaBi: origin of field-induced resistivity upturn and plateau in compensated semimetals. New Journal of Physics* **18**, 082002(2016).
25. Tafti, F.F., Gibson, Q.D., Kushwaha, S.K., Haldolaarachchige, N. and Cava, R.J.: *Resistivity plateau and extreme magnetoresistance in LaSb. Nature Physics* **12**, 272(2016).
26. Arajs, S. and Colvin, R.V.: *Electrical resistivity due to electron-phonon scattering in yttrium and lutetium. Journal of the Less Common Metals* **4**, 572(1962).
27. Destraz, D., Das, L., Tsirkin, S.S., Xu, Y., Neupert, T., Chang, J., Schilling, A., Grushin, A.G., Kohlbrecher, J., Keller, L., and Puphal, P.: *Magnetism and anomalous transport in the Weyl semimetal PrAlGe: possible route to axial gauge fields. npj Quantum Materials* **5**, 1(2020).
28. Destraz, D., Ilin, K., Siegel, M., Schilling, A. and Chang, J.: *Superconducting fluctuations in a thin NbN film probed by the Hall effect. Physical Review B* **95**, 224501(2017).
29. Momma, K. and Izumi, F.: *VESTA 3 for three-dimensional visualization of crystal, volumetric and morphology data. Journal of applied crystallography* **44**, 1272(2011).